\documentclass[onecolumn]{aastex63}
\usepackage{amsmath}
\begin{document}

\title{Black Hole Mass Function of Coalescing Binary Black Hole Systems: Is There a Pulsational Pair Instability Mass Cutoff?}
\author{Yuan-Zhu Wang}
\affil{Key Laboratory of Dark Matter and Space Astronomy, Purple Mountain Observatory, Chinese Academy of Sciences, Nanjing, 210033, People's Republic of China.}
\author{Shao-Peng Tang}
\affil{Key Laboratory of Dark Matter and Space Astronomy, Purple Mountain Observatory, Chinese Academy of Sciences, Nanjing, 210033, People's Republic of China.}
\affil{School of Astronomy and Space Science, University of Science and Technology of China, Hefei, Anhui 230026, People's Republic of China.}
\author{Yun-Feng Liang}
\affil{Guangxi Key Laboratory for the Relativistic Astrophysics, Department of Physics, Guangxi University, Nanning 530004, People's Republic of China.}
\author{Ming-Zhe Han}
\affil{Key Laboratory of Dark Matter and Space Astronomy, Purple Mountain Observatory, Chinese Academy of Sciences, Nanjing, 210033, People's Republic of China.}
\affil{School of Astronomy and Space Science, University of Science and Technology of China, Hefei, Anhui 230026, People's Republic of China.}
\author{Xiang Li}
\author{Zhi-Ping Jin}
\author{Yi-Zhong Fan}
\author{Da-Ming Wei}
\email{Corresponding authors:~yzfan@pmo.ac.cn (YZF) and dmwei@pmo.ac.cn (DMW)}
\affil{Key Laboratory of Dark Matter and Space Astronomy, Purple Mountain Observatory, Chinese Academy of Sciences, Nanjing, 210033, People's Republic of China.}
\affil{School of Astronomy and Space Science, University of Science and Technology of China, Hefei, Anhui 230026, People's Republic of China.}

\begin{abstract}
We analyze the LIGO/Virgo GWTC-2 catalog to study the primary mass distribution of the merging black holes. We perform hierarchical Bayesian analysis, and examine whether the mass distribution has a sharp cutoff for primary black hole masses below $65 M_\odot$, as predicted in pulsational pair instability supernova model. We construct two empirical mass functions. One is a piece-wise function with two power-law segments jointed by a sudden drop. The other consists of a main truncated power-law component, a Gaussian component, and a third very massive component. Both models can reasonably fit the data and a sharp drop of the mass distribution is found at $\sim 50M_\odot$, suggesting that the majority of the observed black holes can be explained by the stellar evolution scenarios in which the pulsational pair-instability process takes place. On the other hand, the very massive sub-population, which accounts for at most several percents of the total, may be formed through hierarchical mergers or other processes.
\end{abstract} 

\section{Introduction}\label{sec:intro}
The successful detection of a gravitational wave (GW) signal from the merger of a binary black hole (BBH) by Advanced LIGO \citep[aLIGO;][]{2016PhRvL.116f1102A} on September 14, 2015, opened a brand-new window into observing the Universe. So far, three observing runs (O1-O3) have been finished by LIGO and Virgo, and the data of several tens of events are released to the public, including two confirmed binary neutron star (BNS) events, 44 confident BBH events, and one neutron star$-$black hole (NSBH) candidate \citep[][see however \citet{2020ApJ...891L...5H} for a dedicated investigation on the possible NSBH nature of GW190425 \citep{2020ApJ...892L...3A}]{2020arXiv201014527A, 2020arXiv201014533T}. 

In population studies, the black hole mass functions (BHMFs) in different binary systems are one of the key subjects since such information can help us reveal the stellar evolution physics and the origin of these systems. The BHMF of coalescing binary black hole systems can not be tightly constrained with the events detected in the O1 run of the Advanced LIGO. Assuming a power-law BHMF with an exponential cutoff at the mass of $\sim 40M_\odot$, \citet{2017arXiv170501881L} predicted that the birth of the lightest intermediate mass black hole (LIMBH, which has a final mass of $\geq 100\,M_\odot$) is very likely to be caught by the Advanced LIGO/Virgo detectors in their O3 run. The data of O1 and O2 observing runs (see \citet{2019PhRvX...9c1040A} for the first Gravitational-Wave Transient Catalog (GWTC-1)), however, strongly favor an abrupt cutoff of the BHMF much sharper than the exponential one. Such a sharp cutoff in the mass spectrum is anticipated in the pulsational pair-instability supernovae (PPISN) scenarios \citep{1964ApJS....9..201F, Belczynski2016, 2020ApJ...896...56W}. In this case, the pre-merger BH can not have a mass substantially more massive than $\sim 40 M_\odot$ and no LIMBHs formed in the BBH mergers are expected in the O3 run. However, some BBHs observed in O3a have masses much larger than those in GWTC-1. In particular, even the secondary mass of GW190521 is more massive than $45 M_\odot$ at $90\%$ credibility and an LIMBH was formed in this event \citep{2020ApJ...900L..13A}. \citet{2020arXiv201014533T} further showed that the simple truncated power-law model is disfavored compared with other more complected models, and the mass spectrum must extend to masses much higher than $40 M_\odot$. The collaboration also claimed the detections of non-zero $\chi_{\rm p}$ and anti-aligned spins in the population. These new discoveries might indicate the presence of multiple formation channels, and some studies were carried out in order to investigate the origins and branch ratios of the sub-populations of BBHs \citep{2020arXiv201202786H, 2020arXiv201105332K, 2021arXiv210203809D}.

In stellar evolution theories, the exact position for the lower edge of the mass gap produced by PPISN depends on the details of the evolution models. Previous studies typically placed the lower edge of the gap at $\leq 65M$ \citep{Belczynski2016,2017MNRAS.470.4739S,2017ApJ...836..244W,2019ApJ...882..121S,2020ApJ...888...76M}. After the discovery of GW190521, some recent works addressed that the gap is most sensitive to the $^{12}{\rm C}(\alpha,\gamma)\,^{16}{\rm O}$ reaction rate, and it could be also affected by other factors like the evolution of H-rich envelope \citep{2019ApJ...887...53F, 2021MNRAS.501.4514C}. By adopting low $^{12}{\rm C}(\alpha,\gamma)\,^{16}{\rm O}$ rates, the maximum black hole mass below the gap can reach $\sim 90 M_\odot$ \citep{2019ApJ...887...53F}, or the mass gap is completely removed \citep{2021MNRAS.501.4514C}. 

PPISNe and pair instability supernovae (PISNe) could affect all formation channels in which the BHs are of stellar origin, including but not limited to classic isolated binary evolution, dynamical capture in different environments, and formation in triple/quadruple systems (see \citet{2019ApJ...882..121S} and the references therein), so the presence of mass gaps are expected in these channels. BBHs formed from dense stellar environments, hierarchical mergers or primordial black holes could populate the mass gap, however, the branch ratios of these channels are generally considered to be small. Under such a recognition, in this work we assume that: (1) The majority of the observed BBHs are originated from stellar evolution, and their distribution can be described by a truncated power law; (2) The position of the hard cutoff of the power law is limited by the traditional prediction on the PPISN/PISN gap, i.e., $\leq 65 M_\odot$. Our main goal is to construct simple models that can be directly compared to the most preferred models in \citet{2020arXiv201014533T} under the framework of Bayesian inference (i.e., {\it to clarify whether the presence of a sharp cutoff in the mass function at the mass $\leq 65M_\odot$ is still consistent with the data).} The other purpose is to examine whether the data of GWTC-1 events and those obtained in O3a run can be reasonably interpreted within the same mass function model.

The rest of the paper is structured as follows: in Sec.\ref{sec:methods}, we introduce the models for parameterizing the mass distributions, the likelihood of hierarchical inference, and the selection effects. We report our results and make comparison between models in Sec.\ref{sec:mc}, and discuss about constraints on the sub-populations in Sec.\ref{sec:Components}. Sec.\ref{sec:dis} is our Conclusion and Discussion.

\section{Data, Models and Selection Effects}\label{sec:methods}
\subsection{Data Selection}\label{sec:data}
In this population study, we include 34 BBH ($m_1 > m_2 > 3 M_{\odot}$) events observed in O3a with FAR $< 1 \rm{yr}^{-1}$, as well as the 10 BBHs presented in GWTC-1. This choice of events is in the same way as for \citet{2020arXiv201014533T}. To test the robustness of our results, the analysis is performed using different subsets of the data: i.e., all of the selected events in GWTC-2, the 34 events detected in O3a alone, and the 10 events in GWTC-1 alone.

The samples from parameter estimation (PE) of individual event are taken from LIGO Public Document Database (see Sec.\ref{subsec:llh} for details), and for each event, we use 1000 random draws of the primary and secondary mass pairs ($m_1,m_2$) from the PE samples in the analysis. The detector frame component masses in the PE samples are transferred to the masses in source frame using ``Planck15" cosmology in {\it Astropy}.

\subsection{Parameterized Mass Spectra}\label{sec:models}
In \citet{2020arXiv201014533T}, the authors concluded that the primary mass distribution is more consistent with a broken power law, or a power law with a Gaussian feature. We also include these two models in our work to compare their preferences by the data with other models. For the BROKEN POWER LAW model (hereafter Model I), we keep its definition and parameter names identical to the descriptions in \citet{2020arXiv201014533T}. For the POWER LAW $+$ PEAK model (hereafter Model II), we modify the formula of primary mass distribution to
\begin{equation}\label{eq:MC}
\pi(m_1 \mid \lambda, \alpha, m_{\rm min}, \delta_m, m_{\rm max}, \mu_m, \sigma_m) = (1-\lambda)\mathcal{P}'(m_1 \mid \alpha,m_{\rm min}, \delta_m, m_{\rm max}) + \lambda \mathcal{G}'(m_1 \mid m_{\rm min}, \delta_m, \mu_m, \sigma_m),
\end{equation}
with
\begin{equation}\label{eq:B}
\mathcal{P}'(m_1 \mid \alpha,m_{\rm min}, \delta_m, m_{\rm max}) = A_1\mathcal{P}(m_1 \mid \alpha,m_{\rm min}, m_{\rm max})S(m_1 \mid m_{\rm min},\delta_{\rm m})
\end{equation} 
being a truncated power-law distribution $\mathcal{P}$ (with spectral index $\alpha$, minimum mass $m_{\rm min}$, and maximum mass $m_{\rm max}$) modulated by a smooth function $S$ (see \citet{2020arXiv201014533T} for details), and 
\begin{equation}\label{eq:G}
\mathcal{G}'(m_1 \mid m_{\rm min}, \delta_m, \mu_m, \sigma_m) = A_2\mathcal{G}(m_1 \mid \mu_m, \sigma_m)S(m_1 \mid m_{\rm min},\delta_{\rm m})
\end{equation} 
being a Gaussian distribution (with mean $\mu_{\rm m}$ and standard deviation $\sigma_{\rm m}$) modulated by the smooth function. $A_1$ and $A_2$ are constants to normalize the distributions, and their values are calculated numerically according to the model parameters. The difference between our expression of Model II and Eq.(B5) of \citet{2020arXiv201014533T} is that both $\mathcal{P}$ and $\mathcal{G}$ are firstly multiplied by the smooth term $S$ {\it before} the normalization. The purpose of making this modification is to ensure that we reproduce Fig.1 of \citet{2018ApJ...856..173T} (the original work that propose this model) giving the same parameter values in that article. The priors on the parameters for Model I and Model II in this work are identical to the priors for the BROKEN POWER LAW model and the POWER LAW $+$ PEAK model in \citet{2020arXiv201014533T}, respectively.

\begin{figure}[]
	\figurenum{1}\label{fig:the2models}
	\centering
	\includegraphics[angle=0,scale=0.5]{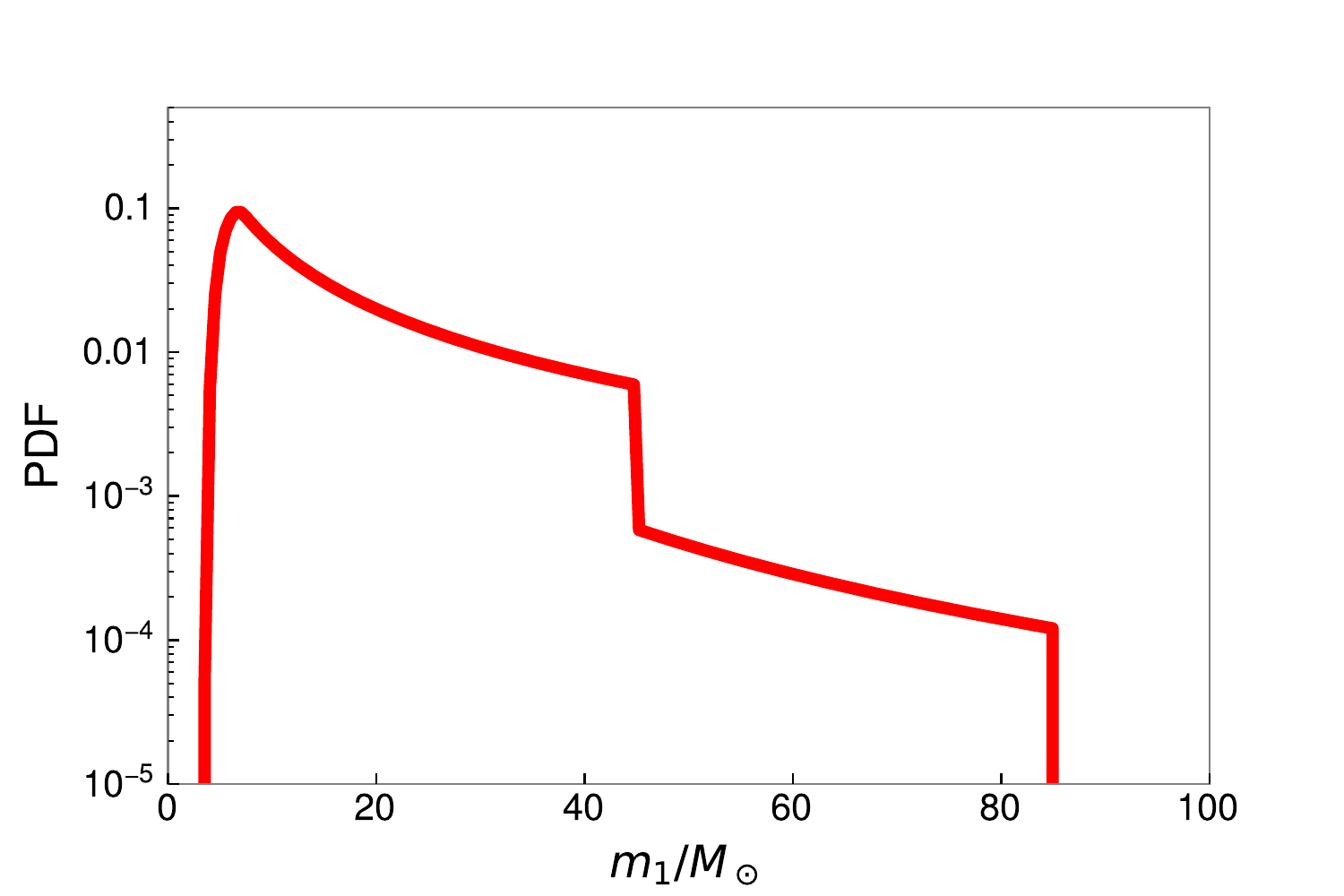}
	\includegraphics[angle=0,scale=0.5]{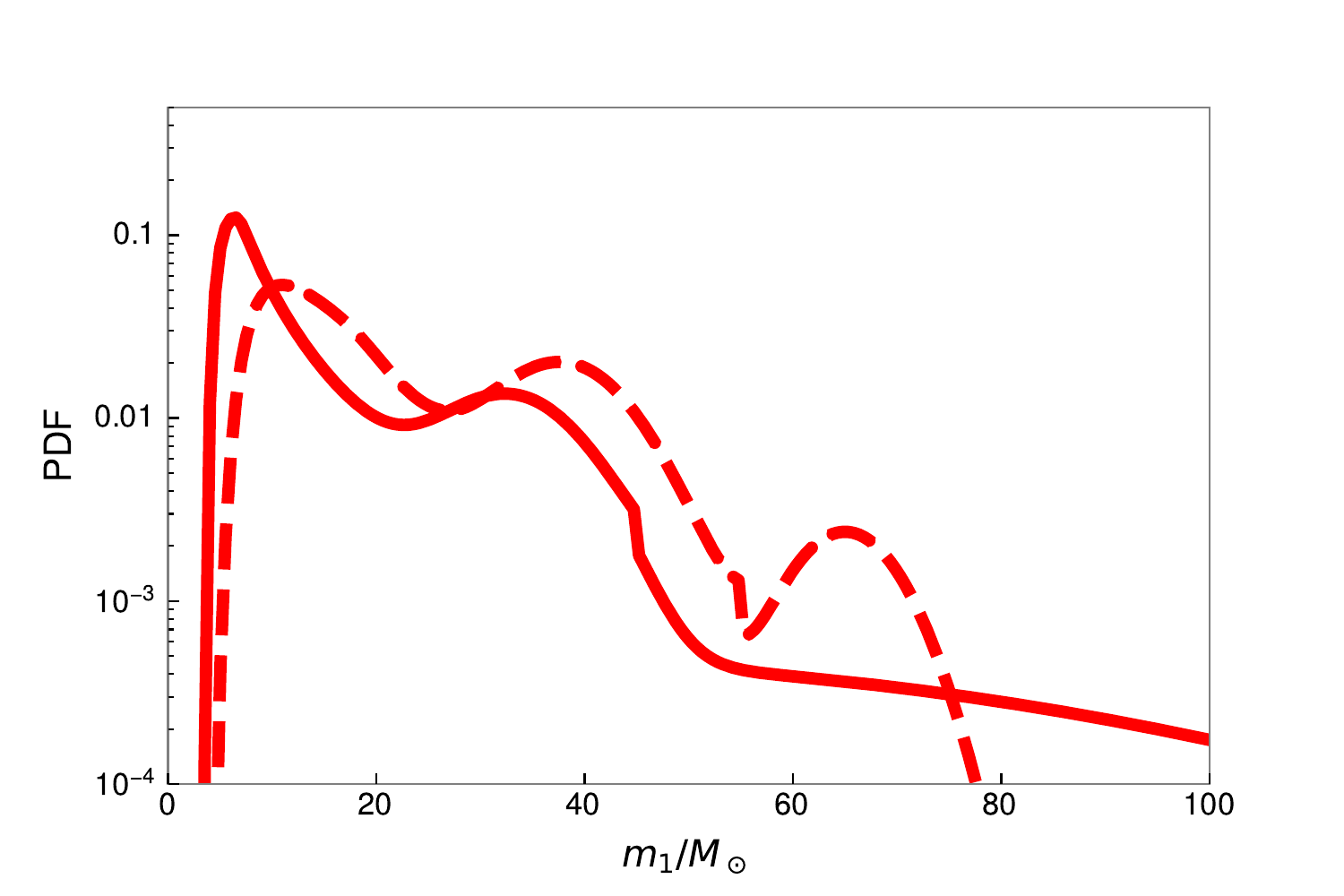}
	\caption{Primary mass functions of Model III and Model IV, for illustration purpose only. Left panel: a representative case for Model III with $F=0.1$; right panel: representatives for Model IV, the solid line shows a case with $\mu_2 \ll m_{\rm max}$ and $\sigma_2 \gg 10$, while the dashed lines shows a case with $\mu_2 \gg m_{\rm max}$ and $\sigma_2 \ll 10$.}
	\hfill
\end{figure}

As mentioned in Sec.\ref{sec:intro}, the stellar evolution scenarios typically predict hard cutoff $< 65 M_{\odot}$. If such a cutoff exists, other evolution channels are needed to explain events with higher masses, and the shape/magnitude of the overall primary mass distribution might change significantly after the cutoff. We first consider a relatively simple case, in which the mass distributions before and after the cutoff are shaped by the power-laws, but with different spectra indices and magnitudes. The corresponding formula for this case (hereafter Model III) is a piece-wise function with two segments,
\begin{equation}\label{eq:M2pl}
\pi (m_1 \mid m_{\rm min},m_{\rm max},m_{\rm edge},\alpha_1,\alpha_2,F,\delta_{\rm m}) = A\times S(m_1 \mid m_{\rm min},\delta_{\rm m}) \times \begin{cases} m_1^{-\alpha_1} & m_{\rm min} \leq m_1 \leq m_{\rm max}
\\ m_1^{-\alpha_2}\, m_{\rm max}^{\alpha_2-\alpha_1}\,F & m_{\rm max} < m_1 \leq m_{\rm edge} \end{cases} ,
\end{equation}
where $\alpha_1$ and $\alpha_2$ are the power law indices for the the segments before and after the cutoff mass $m_{\rm max}$ at the lower edge of mass gap, respectively. $F$ represents the ratio between the possibility densities of the two segments at $m_{\rm max}$. Motivated by the typical predictions about the mass gap, we restrict the prior on $m_{\rm max}$ to be uniform with a maximum of $65 M_\odot$. We expect the second segment to have probability density much smaller than the first segment, so we adopt a log-uniform prior for $F$. To illustrate Model III, we present the representative distribution of the primary mass with arbitrary choice of parameter values in Fig.\ref{fig:the2models}. This model is motivated by some astrophysical theories in which the merging black holes can have different origins. For instance, some BBH systems, in particular those residing within the accretion disks of the Active Galactic Nuclei (AGN) \citep{2017ApJ...835..165B, 2017MNRAS.464..946S, 2018ApJ...866...66M} may be able to accrete material from the surrounding and have higher masses. It has also been proposed that heavy black holes can be dynamically formed by lighter objects through hierarchical mergers or through runaway collisions \citep{Rodriguez2015,2016ApJ...831..187A, 2016MNRAS.459.3432M, 2017ApJ...840L..24F}. For example, the BBH system of GW170729 may be formed through hierarchical mergers in the migration traps that developed in the accretion disks of AGN \citep{2019PhRvL.123r1101Y}. So it is reasonable to expect an extended tail of the mass distribution of the BHs or a population of ``high" mass objects following a sudden drop of the BHMF at $m_{\rm max}$.

Both \citet{2019ApJ...882L..24A} and \citet{2020arXiv201014533T} showed that an extra Gaussian component peaking at $\sim 30-40 M_{\odot}$ might exist. We introduce Model IV, a modified version of the MULTI PEAK model in \citet{2020arXiv201014533T}, to partially study the existence of different components and their influence to the overall shape of the spectrum. Model IV is expressed as
\begin{equation}\label{eq:plgg}
\begin{split}
\pi(m_1 \mid \lambda, \lambda_1, \alpha, m_{\rm min}, \delta_m, m_{\rm max}, \mu_1, \sigma_1, \mu_2, \sigma_2) = (1-\lambda)\mathcal{P}'(m_1 \mid \alpha,m_{\rm min}, \delta_m, m_{\rm max})\\ + \lambda\lambda_1 \mathcal{G}'(m_1 \mid m_{\rm min}, \delta_m, \mu_1, \sigma_1) + \lambda(1-\lambda_1) \mathcal{G}'(m_1 \mid m_{\rm max},\mu_2,\sigma_2),
\end{split}
\end{equation}
where $\mathcal{P}'$ and $\mathcal{G}'$ are described in Eq.(\ref{eq:B}) and Eq.(\ref{eq:G}), (1-$\lambda$) is the fraction of binaries in the main truncated power-law component, and $\lambda\lambda_1$ is the fraction of binaries in the first modulated Gaussian, respectively. For Model IV, we also adopt the astrophysically motivated prior in which $m_{\rm max}$ is uniformly distributed below $65 M_{\odot}$. In addition, comparing with the original MULTI PEAK model, in Model IV the lower bound for the prior on $\mu_2$ is set to $30 M_\odot$, while the upper bound for the prior on $\sigma_2$ is set to $50 M_\odot$. These changes on the prior for the secondary Gaussian component increase its flexibility on describing the spectrum at high masses. The right panel of Fig.\ref{fig:the2models} shows two extreme cases of Model IV: when $\mu_2 \ll m_{\rm max}$ and $\sigma_2 \gg 10$, the second modulated Gaussian actually represents a very shallow component that extends to high masses, while if $\mu_2 \gg m_{\rm max}$ and $\sigma_2 \ll 10$, it represents a clear and narrow Gaussian component after $m_{\rm max}$. 

For all of the four models described above, we use a conditional mass ratio ($q$) distribution that is consistent with Eq.(B8) of \citet{2020arXiv201014533T}, leading to the inclusion of an additional free parameter $\beta_{\rm q}$ in our inference. While different components in the models may have diverse mass ratio distributions, our work mainly focuses on the primary mass distribution, so we leave this issue to future studies. We summarize the parameters of Model I-IV, as well as their priors in Tab.\ref{tb:priors}.

\begin{table*}[]
	\centering
	\caption{Priors of the Parameters for Different Models}
	\begin{ruledtabular}
		\label{tb:priors}
		\begin{tabular}{c|cccc}
			Parameters/Models                                 & Model I    & Model II   & Model III  & Model IV        \\ \hline
			${\rm log_{10}}R_0$[${\rm Gpc^{-3}yr^{-1}}$]      & [-20, 20]  & [-20, 20]  & [-20, 20]  & [-20, 20]       \\
			$\alpha$ or $\alpha_1$                            & [-4, 12]   & [-4, 12]   & [-4, 12]   & [-4, 12]        \\
			$m_{\rm max}$ [$M_\odot$]                         & [30, 100]  & [30, 100]  & [30, 65]   & [30, 65]        \\
			$m_{\rm min}$ [$M_\odot$]                         & [2, 10]    & [2, 10]    & [2, 10]    & [2, 10]         \\
			$\beta_{\rm q}$                                   & [-4, 12]   & [-4, 12]   & [-4, 12]   & [-4, 12]        \\
			$\delta_{\rm m}$ [$M_\odot$]                      & [0, 10]    & [0, 10]    & [0, 10]    & [0, 10]         \\
			$b$                                               & [0,1]      & N/A        & N/A        & N/A             \\
			$\lambda$                                         & N/A        & [0, 1]     & N/A        & [0, 1]          \\
			$\mu_{\rm m}$ or $\mu_1$ [$M_\odot$]              & N/A        & [20, 50]   & N/A        & [20, 50]        \\
			$\sigma_{\rm m}$ or $\sigma_1$ [$M_\odot$]        & N/A        & [0.4, 10]  & N/A        & [0.4, 10]       \\
			$\alpha_2$                                        & [-4, 12]   & N/A        & [-4, 12]   & N/A             \\
			$m_{\rm edge}$ [$M_\odot$]                        & N/A        & N/A        & [65, 100]  & N/A             \\
			${\rm log_{10}}F$                                 & N/A        & N/A        & [-4, 0]    & N/A             \\
			$\lambda_1$                                       & N/A        & N/A        & N/A        & [0, 1]          \\
			$\mu_2$ [$M_\odot$]                               & N/A        & N/A        & N/A        & [30, 100]       \\
			$\sigma_2$ [$M_\odot$]                            & N/A        & N/A        & N/A        & [0.4, 50]       \\
		\end{tabular}
		\tablenotetext{}{The priors on the parameters listed above are all uniform}
	\end{ruledtabular}
\end{table*}

\subsection{The Likelihood and Selection Effects}\label{subsec:llh}
The likelihood for hierarchical Bayesian inference is constructed based on Poisson process. For a series of measurements of $N_{\rm obs}$ events $\vec{d}$, assuming a non-evolving merger rate $R_0$, the likelihood for the hyper-parameters $\Lambda$ (including $R_0$) can be inferred via \citep{2019PASA...36...10T, 2020arXiv201014533T}
\begin{equation}\label{eq:llh}
\mathcal{L}(\vec{d}\mid \Lambda) = N^{N_{\rm obs}}\exp(-N \eta(\lambda))\prod_{i}^{N_{\rm obs}}\frac{\mathcal{Z}_{\varnothing}(d_i)}{n_i}\sum_{k}^{n_i}\frac{\pi(\theta_{i}^k\mid \Lambda)}{\pi(\theta_{i}^k\mid \varnothing)},
\end{equation}
where $N = R_0 V_0 T_{\rm obs}$ is the expected number of mergers during the observation period $T_{\rm obs}$ and within the astrophysical volume $V_0$. Here we take $V_0T_{\rm obs} = 167.6\,{\rm Gpc^{-3} yr^{-1}}$ for O3 and $154.5\,{\rm Gpc^{-3} yr^{-1}}$ for O1-O2. In Eq.(\ref{eq:llh}), the $n_i$ posterior samples for the $i$-th event, the evidence $\mathcal{Z}_{\varnothing}(d_i)$ as well as the default prior $\pi(\theta_k \mid \varnothing)$ are available for both of GWTC-1 \footnote{\url{https://dcc.ligo.org/LIGO-P2000193/public}} and GWTC-2 \footnote{\url{https://dcc.ligo.org/LIGO-P2000223/public}} events. $\eta(\lambda)$ is the detection efficiency for a particular $\lambda$, and we follow the procedures described in the SensitivityTutorial in LIGO Public Document Database to compute this quantity \footnote{The procedures use results of injections, which is from \url{https://dcc.ligo.org/LIGO-P2000434/public} for O1-O2 and \url{https://dcc.ligo.org/LIGO-P2000217/public} for O3}. We use the same criteria that define the detectable events as \citet{2020arXiv201014533T}, i.e., ${\rm SNR} > 8$ for O1-O2 and ${\rm FAR} < 1/{\rm yr}$ for O3. 

Finally, we use the python package {\sc Bilby} and {\sc PyMultinest} sampler to obtain the Bayesian evidence and posteriors of the hyper-parameters for each models.

\section{Models Comparison}\label{sec:mc}
We compute Bayes factors between models to quantify their relative preferences by data. Tab.\ref{tb:bf} shows the Bayes factors $\mathcal{B}$ for each mass model relative to Model II, and in the context we interpret $\mathcal{B}$ of $<1/3$ as moderate, $<1/30$ as strong, and $< 1/100$ as decisive evidence for the first model is less favorable by the data compared with the second model (Model II) \citep{Jeffreys98}. For the analysis including all data, the Bayes factor between Model I and Model II is 0.13, which is consistent with the reported value of 0.12 between the BROKEN POWER LAW model and the POWER LAW$+$PEAK model in \citet{2020arXiv201014533T}. Both Model III and Model IV has larger $\mathcal{B}$ comparing with Model I in the GWTC-2 analysis and O3a-only analysis, while they are less preferred by the data in the GWTC-1 analysis. This result is understandable, since Model III and Model IV are more flexible on describing the high mass spectrum and have larger prior volumes. GWTC-1 contains relatively lighter BHs on average, so Model III and Model IV suffer from Occam factor penalty in Bayes factor; the fraction of BHs with large masses increase significantly in O3a, hence the introducing of larger prior volumes allows the models to fit the data better, giving higher likelihoods; the outcome of GWTC-2 analysis can be regarded as the average over the GWTC-1 and O3a-only analyses. On the other hand, there is no strong evidence ($\mathcal{B}<1/30$) that one of the four models is better supported by the data against others. In general, one conclusion that can be made at this stage is that Model III and Model IV give comparable goodness of fit to the data compared with Model I and Model II respectively, indicating they are also acceptable approximations to the observed population. Since Model III and Model IV are not strongly disfavored by the GWTC-1 data, we conclude that there is no strong tension between the data collected in different runs.

\begin{table} 
	\centering \caption{Bayes factors for each mass model relative to Model II.}
	\label{tb:bf}
	\begin{tabular}{lrrr} 
		\hline              \bf{Models}                     & $\mathcal{B}$(GWTC-2)       & $\mathcal{B}$(O3a only)    & $\mathcal{B}$(GWTC-1 only)  \\
		\hline\hline        I                               & $0.13 \pm 0.03$     & $0.61 \pm 0.12$            & $0.19 \pm 0.03$             \\
		                    II                              & 1                   & 1                          & 1                           \\
		                    III                             & $0.21 \pm 0.05$     & $0.70 \pm 0.14$            & $0.18 \pm 0.03$             \\
		                    IV                              & $1.31 \pm 0.38$     & $2.05 \pm 0.42$            & $0.10 \pm 0.02$             \\
		\hline              III($\alpha_2=1$)               & $0.28 \pm 0.06$     & --                         & --                          \\
		                    IV($\mu_2=40, \sigma_2=44$)     & $1.58 \pm 0.33$     & --        & --               \\
		                    IV($\mu_2=83, \sigma_2=17$)     & $0.27 \pm 0.06$     & --        & --               \\
		                    IV(limited $\lambda\lambda_1$)  & $0.23 \pm 0.05$     & --        & --               \\
		\hline\end{tabular}
	    \tablenotetext{}{\raggedright The Bayes factors are derived from Bayesian evidences $\mathcal{E}$ for the corresponding model in the inference. The numerical uncertainties of the $\rm{ln}\mathcal{E}$ calculated by our nested sampling procedure are propagated onto the Bayes factors.}
\end{table}

\section{Constraints of the potential Components}\label{sec:Components}
\begin{figure}[]
	\figurenum{2}\label{fig:m3}
	\centering
	\includegraphics[angle=0,scale=1.0]{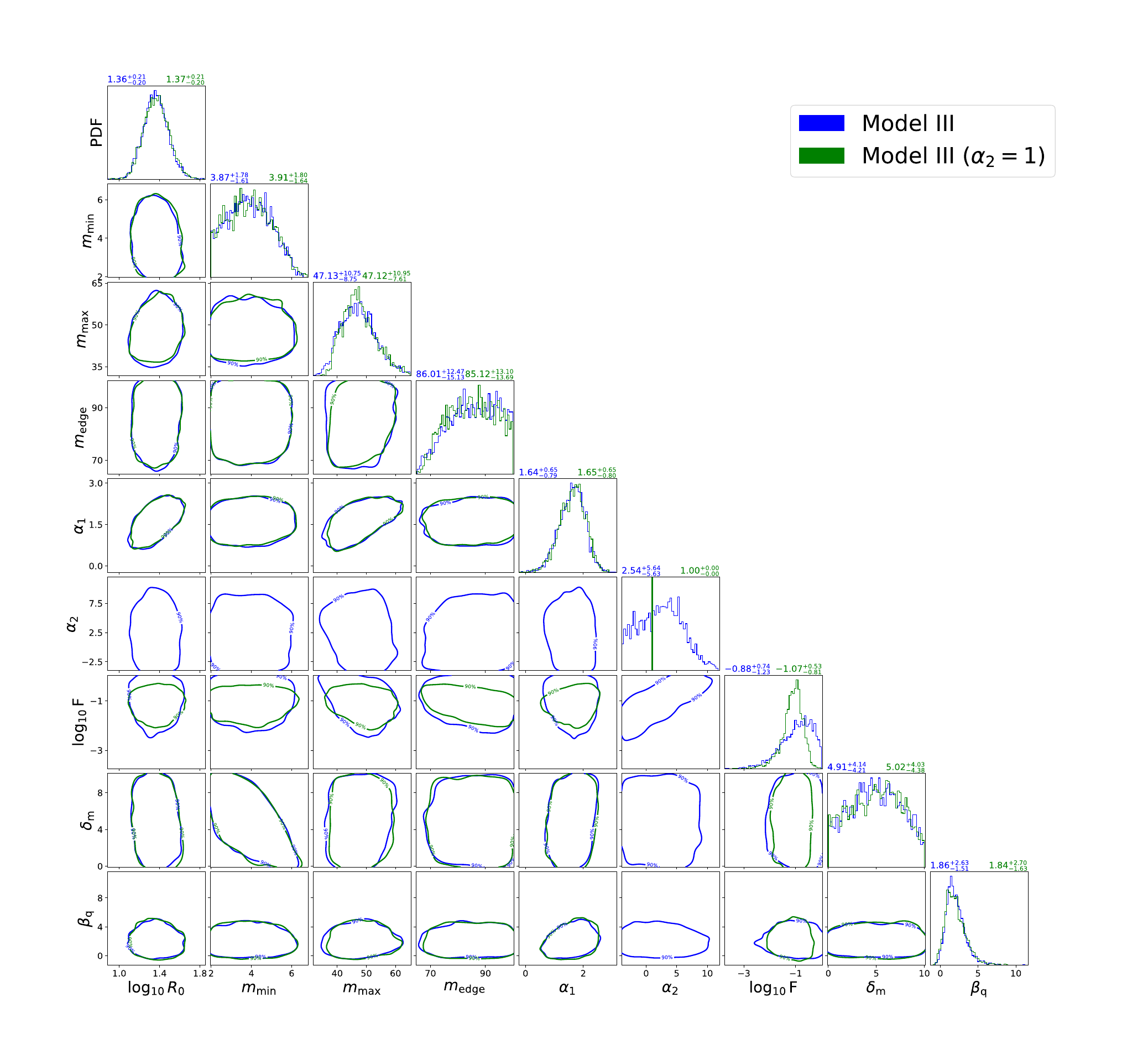}
	\caption{Posterior distributions for the parameters of Model III. The values above the diagonal corner plots represent the 90\% credible intervals.}
	\hfill
\end{figure}

\begin{figure}[]
	\figurenum{3}\label{fig:m4}
	\centering
	\includegraphics[angle=0,scale=1.0]{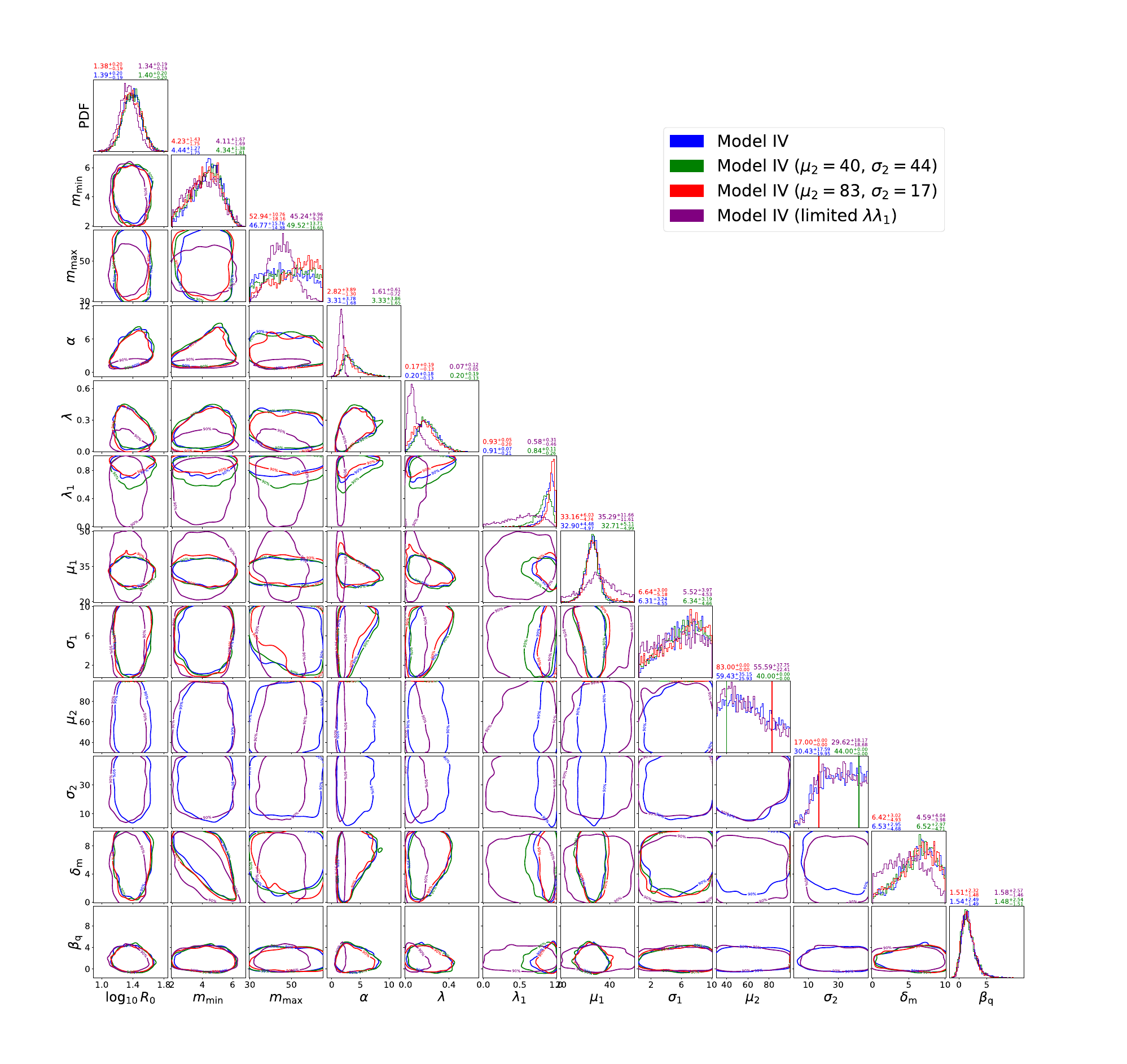}
	\caption{Posterior distributions for the parameters of Model IV. The values above the diagonal corner plots represent the 90\% credible intervals.}
	\hfill
\end{figure}

In this section, we mainly focus on the constraints obtained using all data (the GWTC-2 analysis). To present the constraints on each model, we summarize the median and 90 percent credible intervals of the hyper-parameters for the GWTC-2 analysis in Tab.\ref{tb:constraint}. The posterior distribution for the parameters of Model III and Model IV are also shown in Fig.\ref{fig:m3} and Fig.\ref{fig:m4} respectively. The inferred parameters for Model I and Model II in our work are consistent with the results in \citet{2020arXiv201014533T}, despite the peak of the posterior distribution for $\lambda$ in Model II is shifted to a higher value due to our modification on the formula of POWER LAW $+$ PEAK model (see Eq.\ref{eq:MC} for details). 

\begin{table*}[]
	\centering
	\caption{Summary of Constraints on the Parameters Considered in Tab.\ref{tb:priors}}
	\begin{ruledtabular}
		\label{tb:constraint}
		\begin{tabular}{c|cccc}
			Parameters/Models         & Model I                     & Model II                    & Model III                   & Model IV                    \\ \hline
			${\rm log_{10}}R_0$[${\rm Gpc^{-3}yr^{-1}}$]       & ${1.38}_{-0.19}^{+0.21}$    & ${1.38}_{-0.19}^{+0.20}$    & ${1.36}_{-0.20}^{+0.21}$    & ${1.39}_{-0.19}^{+0.20}$    \\
			$\alpha$ or $\alpha_1$    & ${1.65}_{-0.84}^{+0.78}$    & ${2.86}_{-0.66}^{+0.91}$    & ${1.64}_{-0.79}^{+0.65}$    & ${3.31}_{-1.68}^{+3.78}$    \\
			$m_{\rm max}$ [$M_\odot$] & ${85.30}_{-14.28}^{+12.84}$ & ${85.74}_{-14.06}^{+12.49}$ & ${47.13}_{-8.75}^{+10.75}$  & ${46.77}_{-14.38}^{+15.76}$ \\
			$m_{\rm min}$ [$M_\odot$] & ${3.88}_{-1.61}^{+1.82}$    & ${4.20}_{-1.71}^{+1.61}$    & ${3.87}_{-1.61}^{+1.78}$    & ${4.44}_{-1.75}^{+1.27}$    \\
			$\beta_{\rm q}$           & ${2.03}_{-1.63}^{+3.31}$    & ${1.67}_{-1.54}^{+2.90}$    & ${1.86}_{-1.51}^{+2.63}$    & ${1.54}_{-1.49}^{+2.49}$    \\
			$\delta_{\rm m}$ [$M_\odot$]         & ${4.95}_{-4.29}^{+4.20}$    & ${6.16}_{-4.80}^{+3.29}$    & ${4.91}_{-4.21}^{+4.14}$    & ${6.53}_{-4.68}^{+2.95}$    \\
			$b$                       & ${0.45}_{-0.12}^{+0.27}$    & --                          & --                          & --                          \\
			$\lambda$                 & --                          & ${0.17}_{-0.09}^{+0.16}$    & --                          & ${0.20}_{-0.13}^{+0.18}$    \\
			$\mu_{\rm m}$ or $\mu_1$ [$M_\odot$]       & --                          & ${33.11}_{-4.95}^{+3.27}$   & --                          & ${32.90}_{-4.97}^{+4.48}$   \\
			$\sigma_{\rm m}$ or $\sigma_1$ [$M_\odot$] & --                          & ${5.31}_{-3.80}^{+4.08}$    & --                          & ${6.31}_{-4.55}^{+3.24}$    \\
			$\alpha_2$                & ${6.12}_{-2.51}^{+4.17}$    & --                          & ${2.54}_{-5.63}^{+5.64}$    & --                          \\
			$m_{\rm edge}$ [$M_\odot$]& --                          & --                          & ${86.01}_{-15.13}^{+12.47}$ & --                          \\
			${\rm log_{10}}F$         & --                          & --                          & ${-0.88}_{-1.23}^{+0.74}$   & --                          \\
			$\lambda_1$               & --                          & --                          & --                          & ${0.91}_{-0.21}^{+0.07}$    \\
			$\mu_2$ [$M_\odot$]                   & --                          & --                          & --                          & ${59.43}_{-25.93}^{+35.15}$ \\
			$\sigma_2$ [$M_\odot$]               & --                          & --                          & --                          & ${30.43}_{-19.95}^{+17.59}$ \\
		\end{tabular}
		\tablenotetext{}{\raggedright The table shows the median and $90\%$ credible intervals of posterior distributions, inferred from the analysis using all 44 samples. The analysis performed with only O3a events gives consistent and slightly looser constraints on the parameters.}
	\end{ruledtabular}
\end{table*}

\begin{figure}[]
	\figurenum{4}\label{fig:pdfband}
	\centering
	\includegraphics[angle=0,scale=0.5]{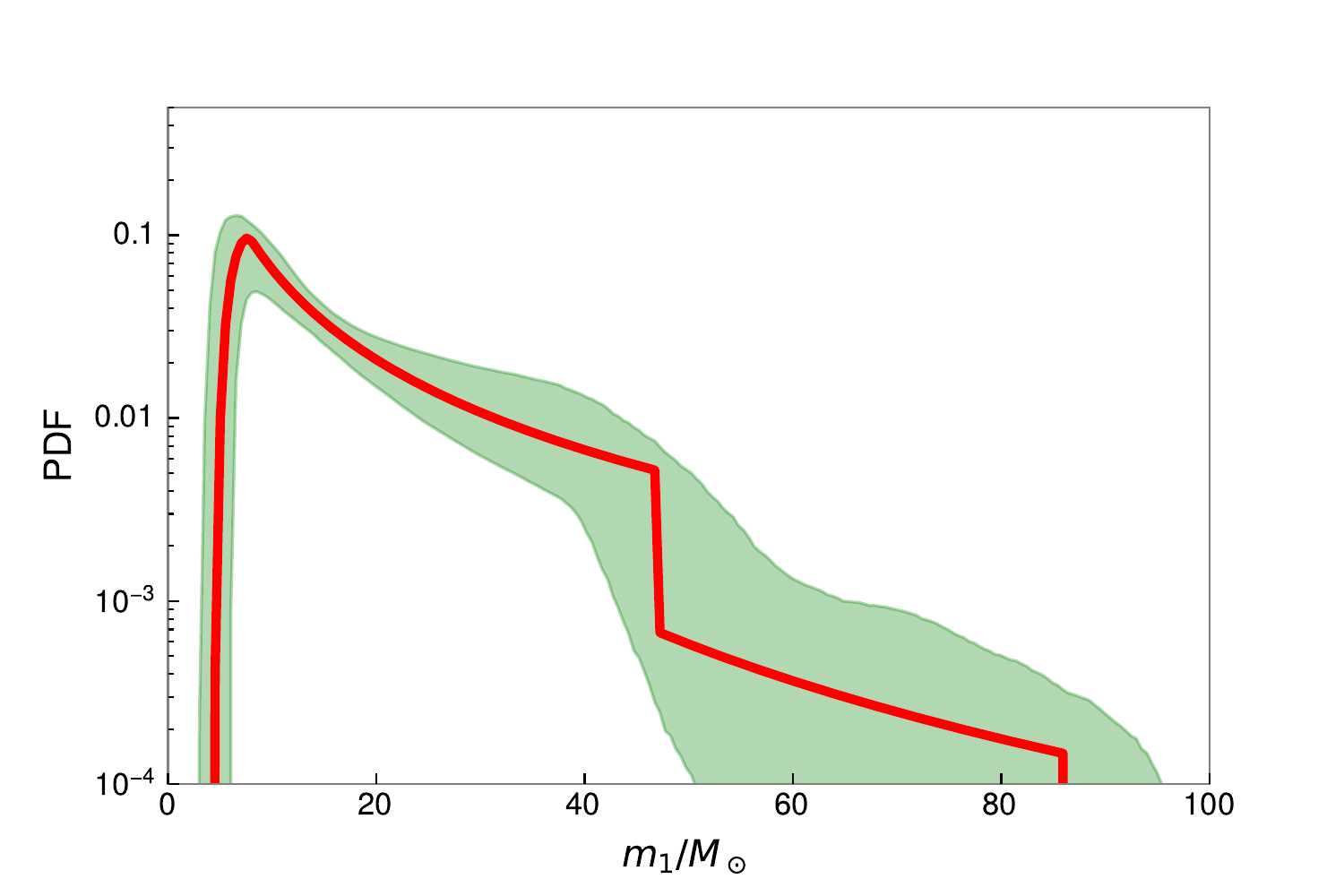}
	\includegraphics[angle=0,scale=0.5]{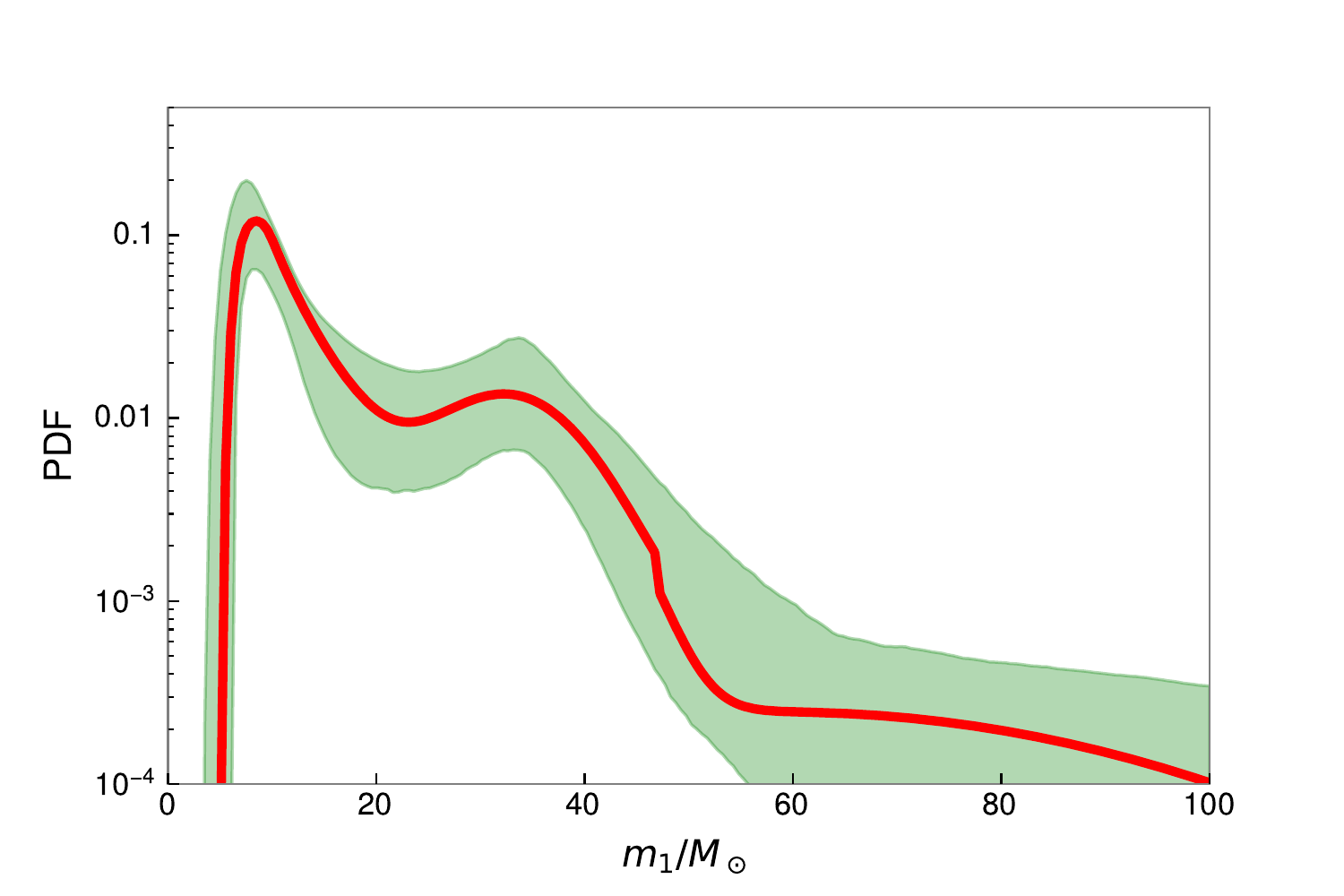}
	\caption{Inferred probability density function (PDF) of astrophysical primary black hole for Model III (left) and Model IV (right). The red lines are the representative distributions drawn by adopting the medians of each parameter's posterior distribution. The green region shows the $90\%$ credible interval derived from the posteriors of the hyper-parameters.}
	\hfill
\end{figure}

For Model III, the inferred parameters that describe the first power-law segment are in agreement with the ones in Model I. The maximum mass ($m_{\rm max}$) of this segment is constrained to ${47.13}_{-8.75}^{+10.75} M_\odot$, above which the probability density function (PDF) of the primary mass spectrum falls by a factor of $0.01-0.72$ (according to the credible interval of ${\rm log_{10}}F$) and the power-law index $\alpha_2$ changes to ${2.54}_{-5.63}^{+5.64}$. Interestingly, the position of $m_{\rm max}$, as well as the first power-law index $\alpha_1$ of Model III is also in agreement with the inferred $\alpha$ and $m_{\rm max}$ for the truncated power law model in previous study (see the results for Model B in \citet{2019ApJ...882L..24A}) using GWTC-1 data. We also find that the constraint on $m_{\rm max}$ is insensitive to the PPISNe motivated prior of $m_{\rm max} \leq 65 M_{\odot}$. By changing the prior for $m_{\rm max}$ to $30 M_\odot \leq m_{\rm max} \leq 95 M_\odot$, and fixing $M_{\rm edge} = 100 M_{\odot}$, the resulting constraint is $m_{\rm max}={49.15}_{-10.59}^{+22.26} M_\odot$. Together with other constrained parameters, we can infer that $0.1\%-5.8\%$ of the primary BHs have masses larger than $m_{\rm max}$. We show the credible region of the mass spectrum for Model III in the left panel of Fig.\ref{fig:pdfband}. Since $\alpha_2$ is poorly constrained, we further fix it to 1, which represents a very shallow tail after $m_{\rm max}$, and the analysis using all data also gives an acceptable Bayes factor of $\mathcal{B}=0.28$ for Model III. We can see that by introducing an abrupt drop on the mass spectrum at $m_{\rm max}$, Model III allows a much shallower segment at high masses compared with Model I.

\begin{figure}[]
	\figurenum{5}\label{fig:pdfband2}
	\centering
	\includegraphics[angle=0,scale=0.3]{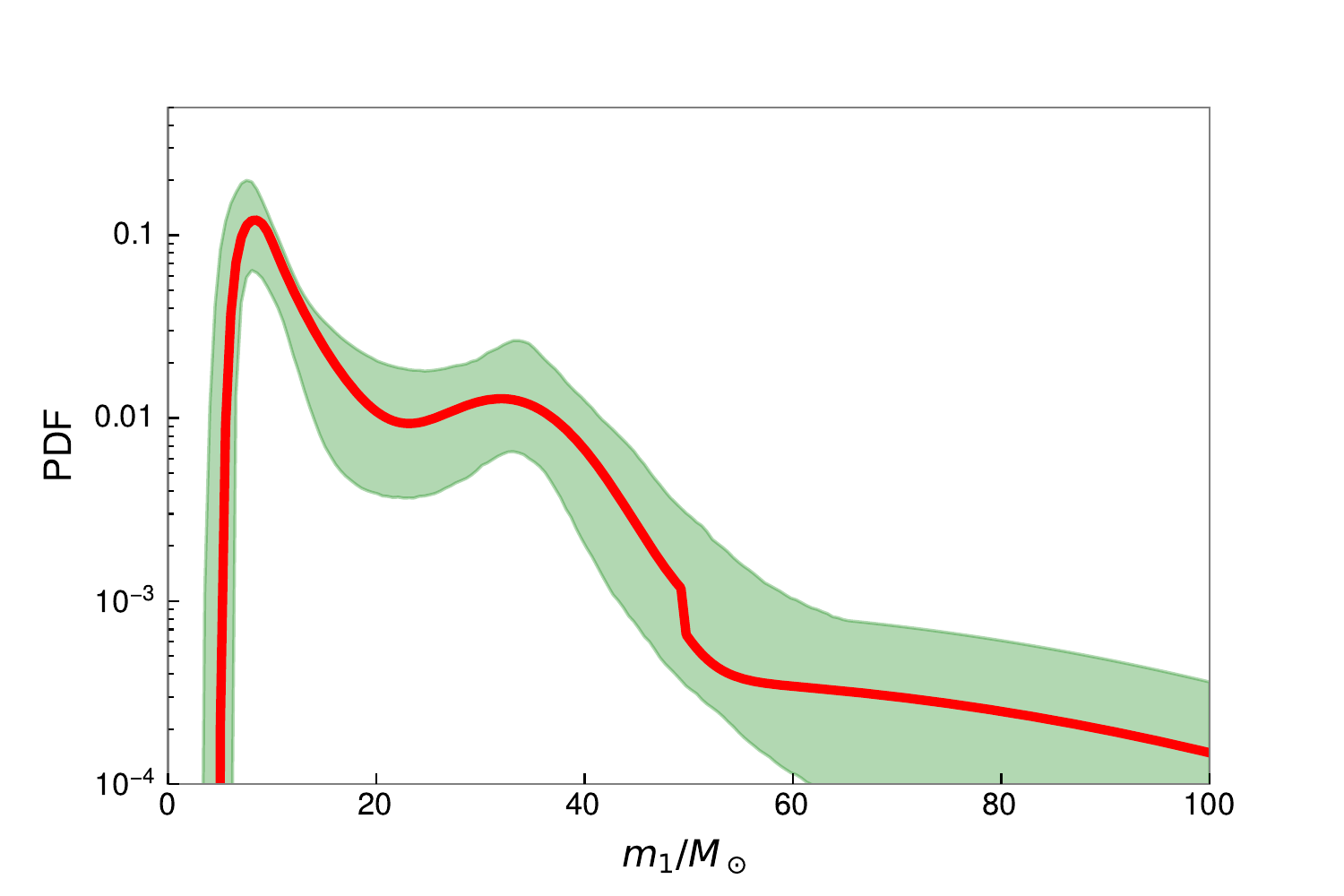}
	\includegraphics[angle=0,scale=0.3]{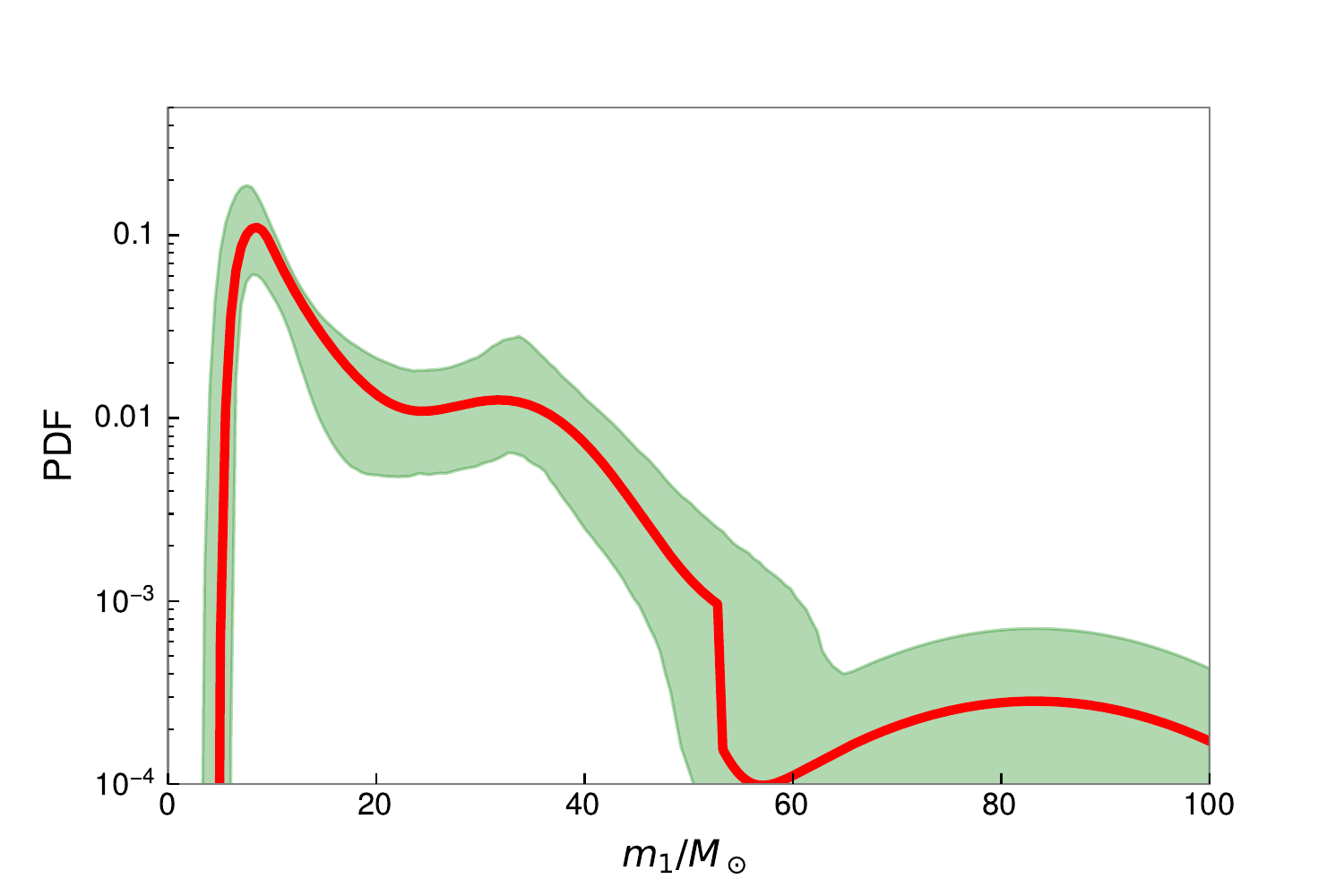}
	\includegraphics[angle=0,scale=0.3]{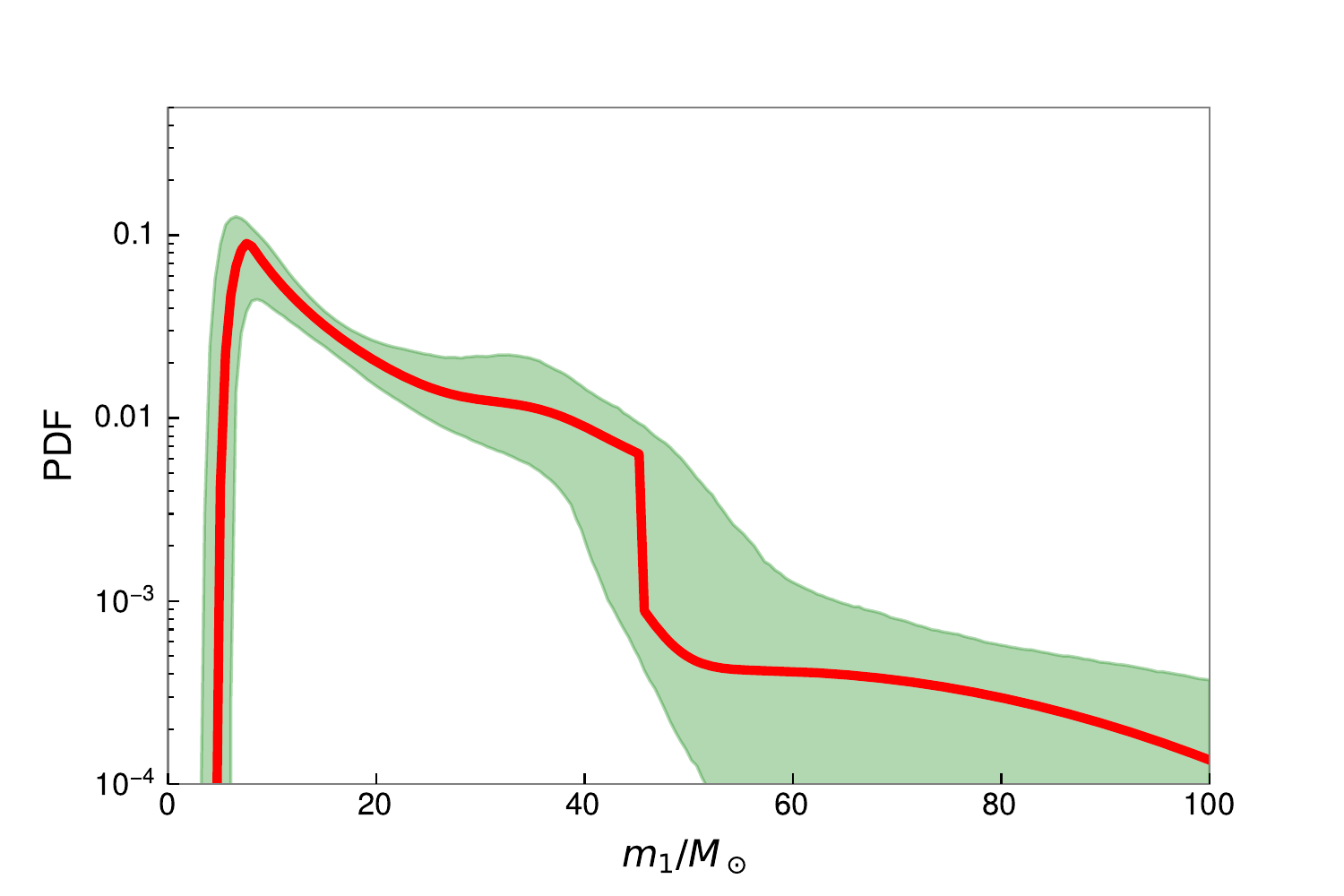}
	\caption{The same as Fig.\ref{fig:pdfband}, but for the three special cases of Model IV. Left panel: the case with $(\mu_2=40,\sigma_2=44)$; middle panel: the case with $(\mu_2=83,\sigma_2=17)$; right panel: the case with additional restriction described in Eq.(\ref{eq:lamlim}).}
	\hfill
\end{figure}

For Model IV, $m_{\rm max}$ is less constrained. As shown in Fig.\ref{fig:m4} the posterior of $m_{\rm max}$ is broadly distributed across the range of prior, and there is significant posterior support around the median of $m_{\rm max}$ inferred from Model III. The result also shows a Gaussian component peaking at $\sim 33 M_\odot$, regardless of whether we use all events or O3a-only events in the analysis. Both the study in \citet{2019ApJ...882L..24A} and the O3a-only analysis in our work have recovered consistent peaks for the Gaussian component, which enhances the evidence about the presence of this sub-population of black holes. The second Gaussian component, which contains $0.4\%-4.9\%$ of all primary BHs, is poorly constrained. From the posterior distributions, we can only exclude small values ($<10 M_\odot$) of $\sigma_2$. The $\mu_2$ for this component can be either smaller than $\mu_1$ of the first Gaussian component, or larger than $m_{\rm max}$. The credible region for Model III is shown in the right panel of Fig\ref{fig:pdfband}, and due to the large uncertainties on the second Gaussian component, it could alternatively represent a weak and shallow component rising below $m_{\rm max}$ and extending to higher masses rather than a real Gaussian one (which is similar to the case marked with dashed lines in Fig.\ref{fig:the2models}). To quantify the preference for the shape of this component by data, we further fix $(\mu_2=83,\sigma_2=17)$ and $(\mu_2=40,\sigma_2=44)$ (chosen according to the $68\%$ upper and lower bounds of the posterior distributions for $\mu_2$ and $\sigma_2$) to reanalyze the data. Posterior distributions for the parameters and the credible regions of the primary mass spectra of these two cases are presented in Fig.\ref{fig:m4} and Fig.\ref{fig:pdfband2}. The resulting Bayes factor between the two cases is $\mathcal{B}=0.17$, which indicates a modest support for the component being a shallow one peaks below $m_{\rm max}$. Nevertheless, it is still lack of strong evidence to exclude the case in which the component has a relatively narrower ($\sigma_2 \lesssim 20$) Gaussian shape and peaks after $m_{\rm max}$. 

Another possible effect of PPISNe in addition to the formation of mass gap is leaving an excess of BHs near the lower edge of the gap \citep{2018ApJ...856..173T}. If this excess fully accounts for the first Gaussian component in Model IV, the number of black holes in such component should be no more than the number of black holes that would have been formed by the power-law component continued to the upper limit of the mass gap $m_{\rm PI}$ \citep{2018ApJ...856..173T}, i.e.,
\begin{equation}\label{eq:lamlim}
\lambda \, \lambda_1 \leq \frac{\int_{m_{\rm max}}^{m_{\rm PI}}\mathcal{P}'(m_1 \mid \alpha,m_{\rm min}, \delta_m, m_{\rm PI})}{ \\
                          \int_{m_{\rm min}}^{m_{\rm PI}}\mathcal{P}'(m_1 \mid \alpha,m_{\rm min}, \delta_m, m_{\rm PI})} \\
                      \approx \frac{\int_{m_{\rm max}}^{m_{\rm PI}}m^{-\alpha}}{\int_{m_{\rm min}+\delta_{\rm m}}^{m_{\rm PI}}m^{-\alpha}},
\end{equation}
where we take $m_{\rm PI}=150 M_\odot$. To test if this hypothesis is supported by our posterior, we compute the fraction of black-holes, pertaining to the power-law component of the mixture, with a mass above $m_{\rm max}$. We do this using the posteriors for the parameters of Model IV. We find that only $\sim 13\%$ of the posteriors predict such a fraction larger than $\lambda \, \lambda_1$. We further apply the restriction in Eq.(\ref{eq:lamlim}) to the Bayesian inference of Model IV, and find that the resulting Bayes factor (with respect to Model II) is 0.23. Since the $\mathcal{B}$ for Model IV without this restriction is 1.31, the PPISNe origin of the first Gaussian component is less preferred (but not excluded).

\section{Conclusion and Discussion}\label{sec:dis}
In this work, we study the primary mass distribution of the merging BBHs and focus on probing the presence of a sudden drop of the mass function at the black hole mass of $\leq 65M_\odot$, as predicted in pulsational pair instability supernova model. We construct two empirical mass functions, and by performing Bayesian inference, we find that these two models are still comparable with the most preferred empirical models (i.e., Model I and Model II in this work) found in previous studies and a cutoff of the black hole mass function at $\sim 50M_{\odot}$ is indeed consistent with the data (see Fig.\ref{fig:pdfband}). The very massive sub-population, which accounts for at most several percents of the total merging black holes, may be from hierarchical mergers or other processes. Note that in this work we concentrate on the BBH systems. In the future a reasonably large neutron star$-$black hole merger event sample is expected to be available, with which the mass function of these black holes can be reconstructed \citep{2020ApJ...892...56T} and it would be quite interesting to see whether the black hole mass functions are significantly different among different binary systems.

As already discussed in Sec.\ref{sec:Components}, there is a consistency between the inferred parameters for the first segment in Model III and the ones for Model B in \citet{2019ApJ...882L..24A}. We therefore suspect that \citet{2019ApJ...882L..24A} has already recovered spectral shapes that roughly match the actual primary mass distribution below the cutoff, while missed the sub-population above the cutoff due to the small number of the GWTC-1 events. Based on our analysis about Model IV, we find moderate support for the low mass Gaussian component being not originated from PPISNe. On the other hand, if PPISNe truly account for this excess of BHs, the first Gaussian component should be relatively weak, and as shown by the right panel of Fig.\ref{fig:pdfband2}, the credible region for Model IV in this case is very similar to the credible region for Model III (the right panel of Fig.\ref{fig:pdfband}). It is difficult to place good constrains on all of the parameters in Model IV. As shown in Fig.\ref{fig:m4} and Fig.\ref{fig:pdfband}, the three superimposed components may cover up the cutoff feature of the power-law component. Nevertheless, since their origins are unclear, the two sub-dominant components in model IV might essentially belong to one population peaking at $\sim 33 M_\odot$ and having an extended tail. For the purpose of this work, we do not attempt to recover the exact shape of these components by introducing more assumptions (making more complicated models) with current data.

If the primary mass distribution indeed consists of different sub-populations, evidences may be found elsewhere in addition to the mass spectrum. For example, the inclusion of spin data (although additional considerations are needed to construct the spin model) would enhance the ability for model comparison in the inference. The expected spins for BHs with primary mass above the mass cutoff could be larger and more isotropic if we assume these BBHs are formed by dynamical capture. It is worthy of noting that GW190521 has $\chi_{\rm p} \sim 0.6$ \citep{2020ApJ...900L..13A}, which is larger than the $\chi_{\rm p}$ of other observed events with lower masses. On the other hand, the evidence for anti-aligned spin and the non-zero $\chi_{\rm p}$ of the whole population found in \citet{2020arXiv201014533T} may suggest the presence of different formation channels. However, \citet{2020arXiv201014533T} also pointed out that there is no strong evidence for variation of the spin distribution with mass. More events are needed if we want to take all these possibilities into account to make solid conclusions. 

\section{Acknowledgment}
We thank the referee for very helpful suggestions and M. Fishbach, D. E. Holz, Y. M. Hu, and Y. Qin for their kind help. This work was supported in part by NSFC under grants of No. 11921003, No. 11933010, No. 12073080 and No. 11525313, the Funds for Distinguished Young
Scholars of Jiangsu Province (No. BK20180050), the Chinese Academy of Sciences via the Strategic Priority Research Program (Grant No. XDB23040000), Key Research Program of Frontier Sciences (No. QYZDJ-SSW-SYS024). This research has made use of data and software obtained from the Gravitational Wave Open Science Center (https://www.gwopenscience.org), a service of LIGO Laboratory, the LIGO Scientific Collaboration and the Virgo Collaboration. LIGO is funded by the U.S. National Science Foundation. Virgo is funded by the French Centre National de Recherche Scientifique (CNRS), the Italian Istituto Nazionale della Fisica Nucleare (INFN) and the Dutch Nikhef, with contributions by Polish and Hungarian institutes.

\software{Bilby \citep[version 0.6.9, ascl:1901.011, \url{https://git.ligo.org/lscsoft/bilby/}]{2019ascl.soft01011A}, PyCBC \citep[version 1.13.6, ascl:1805.030, \url{https://github.com/gwastro/pycbc}]{2018ascl.soft05030T}, PyMultiNest \citep[version 2.6, ascl:1606.005, \url{https://github.com/JohannesBuchner/PyMultiNest}]{2016ascl.soft06005B}}

\end{document}